# Switching based Spin Transfer Torque Oscillator with zero-bias field and large tuning-ratio

*Gaurav Gupta[1,2][†], Zhifeng Zhu[2] and Gengchiau Liang[2]*

[1]Spin Devices, Delhi 110006, India

[2]Department of Electrical and Computer Engineering, National University of Singapore, Singapore 117576

We propose a novel concept of obtaining oscillations with frequencies in very-high frequency (VHF) and ultra-high frequency (UHF) bands. A traditional spin torque nano-oscillator (STNO) consists of at least one pinned layer (PL) and one free layer (FL) which precesses in a fixed orbit, defined by a precession angle, which results in magneto-resistance (MR) oscillations. In STNO, with aligned or even orthogonal easy-axis of the magnetic layers and with or without external bias magnetic field, it is not possible to attain full MR swing. The constringed MR swing jeopardizes the extracted output power. Furthermore, the orbit is strongly disturbed by the thermal fluctuations resulting in strong magnetic noise. In stark contrast to the operation principle of a STNO, we theoretically demonstrate, with the practical parameters from the experiments, that with a unidirectional current in a dual asymmetric free-layers (with no pinned layer) based perpendicular magnetic tunnel junction (pMTJ), both of the free layers can attain a complete and out-of-phase self-sustained switching without the aid of any external magnetic field. This design facilitates a switching based spin torque oscillator (SW-STO) with a full MR swing, and hence a larger output power, for stable and more thermally robust free-running oscillations. Furthermore, the integration with the n-type metal-oxide-semiconductor (NMOS) field-effect transistors at 130, 65 and 14 nm node is appraised to expound its effect on the oscillator performance, controllability with DC bias and the design constraints, to demonstrate the viability of the design as a dynamically controllable oscillator for practical on-chip implementation.

## I. INTRODUCTION

On-chip oscillators [1] generate stable periodic oscillations, whose voltage (VCO), current (CCO) or digital (DCO) control serves as an integral component of the phase-locked loop (PLL) and radio-frequency (RF) transceivers. PLL generates the clock signal for the chip, where it can either output the same or multiply or divide the VCO frequency to suit the application. Frequency (*f*) control of the oscillators enables dynamic-frequency-scaling (DFS) scheme [2] for optimizing the low-power modes. Furthermore, as the indispensable blocks in the RF transceiver circuits, oscillators are used for modulation and clock recovery. For these applications, an ideal oscillator requires a large quality factor (Q) (i.e. small phase noise), low power consumption ($P_{IN}$), large output power or voltage swing ($P_O$ or $V_O$), integrability with CMOS, large tunability ratio ($f_{max}/f_{min}$) and small die area occupancy in the desired frequency regime. Extant solutions [1, 3] are based on the ring-oscillators, which have low Q, for sub-GHz and LC-tank oscillators (large Q) for a few GHz range, because the phase noise for the ring oscillators further degrades as the frequency is scaled-up. Furthermore, LC oscillators have small tuning range [1, 4], have large $P_{IN}$, are difficult to integrate in the very-large scale integration (VLSI) chips and occupy large area because of the large inductor and varactor size even in low GHz regime at which custom-designed chips can operate.

To overcome these issues, recently spin-torque nano-oscillators (STNOs) and spin-hall nano-oscillators (SHNO) have been proposed [5] and integrated with CMOS [6]. They are expected to occupy small area (~ area of magnetic-tunnel junction (MTJ)), consume less power and provide ultra-wide-band operation [7, 8]. STNOs are based on the precession of a free-layer (or set of locked [9-11] FLs) in MTJ within a certain orbit determined by a precession angle [5], with a single or a set of fixed reference/pinned layer [12]. This precession in a locked orbit results in an oscillatory change in the magnetization of the free layer (s) with respect to that of the reference layer, which translates into a resistance change of a MTJ (or spin-vale in the case of a metallic spacer), $R_{MTJ}$ and thus an output voltage swing. With the application of a well-controlled and precise bias magnetic field $H_{Applied}$, large Q for STNO has been obtained [13]. Even in another genre of spin-oscillators based on vortex (VSTO), where an oscillating magnetic vortex is generated in thick magnetic layers of large cross-section via current induced spin-transfer torque (STT) and perpendicular $H_{Applied}$ [14-19] large Q has been observed. However, providing a uniform and precisely stable magnetic field $H_{Applied}$ in the integrated circuits [19] is extremely difficult, which therefore, would make it difficult to control the oscillator characteristics. Hence, there has been a general effort in recent years to research solutions that can operate without $H_{Applied}$ [20-22]. To eliminate the external bias magnetic-field $H_{Applied}$ required in all in-plane or all-perpendicular designs, MTJ stacks with orthogonal set of layers [10, 23] have also been investigated. To increase the output voltage swing by increasing the observable fraction of the MR in the oscillations, the combination of multiple in-plane and perpendicular magnetization free and pinned/reference layers (orthogonal layers) in a MTJ has also been promulgated for a very large precession angle. It enables *near* full rotation of one of the free layers. However, the use of orthogonal layers across the spacer (metal like Cu or insulator like MgO) inherently limits the MR of these stacks. Furthermore, recently, locked precessional orbits of dual free layer designs, i.e. without any polarizing layer, for in-plane magnetization (ii-MTJ) and with assistance of $H_{Applied}$ have also been shown to produce stable oscillatory output [24, 25]. Without $H_{Applied}$, the output signal has been shown to be very weak ($< -100$ dBm) [26]. Moreover, for multiple precessional orbits with the resistance oscillations depending on the dynamic angle of the precessions, the effect of the magnetic noise becomes stronger in this design compared to a traditional STNO. Furthermore, sophisticated designs with tilted anisotropy, operating without $H_{Applied}$, with weak output ($< -65$ dBm) [21], and more recently with large $P_O$ (nearly $-26$ dBm at 6.7mA current bias) which need very precise control of $H_{Applied}$ and especially designed microstructure for the magnet [27, 28] in sombrero-shaped MTJ, have been demonstrated. As a consequence of the reliance on the fixed precessional orbit, the magnetic noise resulting from the thermal fluctuations strongly affects the performance of a STNO (for instance see Fig. 1 of Ref. [29] or Fig. 5 of Ref. [30]), by randomizing the phase of the precession. This especially becomes pernicious when MTJ area is down-scaled because the thermal field scales as an inverse root of the magnet volume. These issues greatly limit its applicability in the real applications [7].

Therefore, in this work we propose a simple design of a spintronic oscillator, with no exotic structure or tilted anisotropy constraints, as conceptually illustrated in Fig. 1. The generated oscillations (50-960 MHz) are in VHF (30-300 MHz) and UHF (300-1000 MHz) frequency bands [31] which more generally serve application-specific integrated circuits (ASICs) [2, 32-36]. Furthermore, this frequency range also serves the RF band for LoRa protocol (868 MHz for Europe, 915 MHz for North America and 430 MHz for Asia) [37-39] deployed for low-cost bi-directional secure low-power wide-area-network (LPWAN) wireless communication channel in internet-of-things (IoT), machine-to-machine (M2M), smart city and industrial applications [40]. This is in contrast to STNO which is normally over this range and gets to tens of GHz and thus applied as microwave source/detector and in wireless transceivers [5, 7] for other protocols like Bluetooth and Wi-Fi. The proposed design is based on dual asymmetric free-layers, with no pinned layer, in a perpendicular magnetic tunnel junction (pMTJ) and operates without any bias magnetic field. The design has a large tunability ratio. The importance of a large tunability ratio is specifically explained in section III.C. The proposed device can also deliver a relatively large output power because MR oscillations attain a full theoretical maximum. This in turn would be much larger than for MTJs with magnetic layers with orthogonal easy-axis. To achieve this, the design deploys a different operation principle than STNO. It is based on the switching of both the free-layers with a unidirectional current in a pMTJ structure, where one of the free-layer would have traditionally served as a reference layer and mandated current reversal for switching-back the traditional free-layer. The proposed design is subsequently referred as switching based spin torque oscillator (SW-STO). It is shown that both layers can attain complete and out-of-phase self-sustained switching with a DC electrical bias to produce oscillations which are more thermally robust than for a STNO.

The frequency is shown to scale directly with the MTJ current $I_{MTJ}$. In CCO configuration, the design has tunability of over a decade. This is much superior to STNO [41] and ring oscillators [42] for which even attaining an octave (electronic control for STNO) is a challenge, and thus proposed design is better suited for DFS schemes. CCO would, however, warrant an implementation of the stable current sources/mirrors over a wide operating range. This would heavily penalize the design in terms of power and area. Therefore, simply driving a MTJ with a NMOS in the VCO configuration is also appraised. Therefore, as illustrated in Fig. 2(a), the impact of the integration and the control of the oscillator via foundry calibrated NMOS is also examined across the complementary metal–oxide–semiconductor (CMOS) nodes (130, 65 and 14 nm). Either the gate voltage $V_{GS}$ (suggested method in this work) or the node voltage $V_{DD}$ (akin to the ring oscillator based VCOs) of the cell is tuned, where a cell is referred to a system of a MTJ with a (or set of) driving transistors. It is shown that integrating a transistor has a significant impact on the performance metrics of the oscillator. The performance in general degrades, yet is shown to be better than traditional STNOs in the CCO configuration. Furthermore, we show that the NMOS area dominates the cell-size instead of the MTJ area, and hence, should be the one compared with the traditional ring or LC oscillators to claim the area benefits. Moreover, a larger node because of its larger supply voltage has better current drive and thus an iso-frequency comparison exhibits a smaller channel area and a larger output power for a larger node.

The article is organized as follows. In Sec. II, the methodology of our theoretical simulation of the proposed design is explained. Sec. III discusses the results and is anatomized into four sub-sections. In sub-section III.A, the operation principle and the device physics is expounded, the effect of thermal fluctuation i.e. magnetic noise is considered in III.B. Results for the design driven by a current-source thus operating it in a CCO configuration is presented in III.C, while integration of transistors to operate in a VCO and a DCO configuration is assayed in III.D. This is followed in Sec. IV with the summary and the conclusions of our study.

**II. METHODOLOGY**

The simulation framework illustrated in Fig. 2(b) expounds all mechanisms considered in this work, and the flow of the variables across the framework. Two methods are considered in this work to drive a unidirectional current through a pMTJ based proposed oscillator. The first one is a constant current source (akin to suggested for a STNO) which operates the device in the CCO configuration (Fig. 1), and corresponds to only the MTJ dynamics in the framework. The second one is integrating an NMOS with a MTJ on the drain end of the transistor whose supply node $V_{DD}$ (akin to ring-oscillators) or gate (suggested method for this design) is controlled via input DC voltage, which operates the design in the VCO configuration (Fig. 2(a)). This invokes the full framework shown in Fig. 2(b). Other configurations, like a MTJ on the source end or use of the MOS in the diode mode results in an inferior performance of the oscillator and thus not shown here.

The magnetization dynamics is solved via coupled Landau–Lifshitz–Gilbert (LLG) equation integrated via fourth order Runge-Kutta method [43-45]. An alternate method for solving this stochastic partial differential equation (SPDE) is first-order Euler method [46, 47], and the trade-offs for both the methods are discussed in Refs. [44, 48] for a more interested reader. The dynamics of both the FLs is analyzed via macrospin model which has been shown to be valid in scope of the dimensions considered in this work [49-51]. The problem is numerically solved in Matlab which is dynamically coupled with SMASH to solve at every time-interval $\Delta t$ of 5 ps for an updated angle between the two magnetization vectors and the voltage across the MTJ. To find suitable $\Delta t$, for a few random sets of physical and electronic parameters within the operational regime of the design, frequency of the oscillator was plotted against $\Delta t$ swept from 250 fs to 12.5 ps. The trend was observed to remain saturated from 250 fs to 8 ps after which it rolls-off because of the error induced in the LLG equation due to a large time step. Within this saturated region, the value of 5 ps was chosen because the run time was exactly divisible by this time-step and thus allowed to exactly count the cycles, and furthermore being a coarse enough value enabled us to reduce the computational time and storage without affecting the results. The oscillator characteristics are extracted for 10 μs run which results in two million sampling points of the data. It is subsequently parsed by Fast-Fourier Transform (FFT) to extract the oscillator characteristics like fundamental frequency, linewidth, quality factor and output voltage swing $V_O$. The output power $P_O$ in dBm is computed as *10 + 20 log₁₀(V_O)*. For input-power $P_{IN}$ and area, in a CCO configuration only MTJ is considered (akin to the STNO literature) while in a VCO configuration both MTJ and transistor are accounted. Next, the Resistance-Area product or RA = 4.7 Ω-μm$^2$ is taken from CoFeB pMTJ from imec [52] with zero-bias TMR (Tunneling Magnetoresistance) i.e. $TMR_0$ = 143% and MgO thickness $t_{MgO}$ = 1 nm. Throughout the text, MTJ dimensions (in nm) along x, y and z axis are respectively expressed as: length × width × thickness_$FL_1$ (thickness_$FL_2$), with default value of 50nm×50nm×1.6nm(1.2nm). Both damping and field-like torque (DLT and FLT), $a_\parallel$ and $a_\perp$ respectively, are considered in the coupled LLG equation as [53],

$$\left(\frac{1+\alpha^2}{\gamma}\right)\frac{d\mathbf{M}_i}{dt} = -\mathbf{M}_i \times \left[\mathbf{B}^i_{Eff} - (a^i_\perp + \alpha a^i_\parallel)\mathbf{p}_j\right] - \frac{\mathbf{M}_i}{M_S} \times \left\{\mathbf{M}_i \times \left[\alpha\mathbf{B}^i_{Eff} - (\alpha a^i_\perp - a^i_\parallel)\mathbf{p}_j\right]\right\} \quad (1)$$

where α = 0.01 [53] is the damping constant, γ is the gyromagnetic ratio, i =1 and 2 denotes $FL_1$ and $FL_2$, respectively, and $\mathbf{M}_i$ is the magnetization vector of the $i^{th}$ FL. Corresponding magnetization unit vector $\mathbf{m}^i$ is expressed as $[m^i_x\ m^i_y\ m^i_z]$. $\mathbf{p}_j$ is the unit-vector of the spin-angular momentum acting on the magnetic moments of the $i^{th}$ FL due to the $j^{th}$ FL, where i≠j. Hence, $\mathbf{p}_j$ for the $FL_2$ dynamics is parallel to $\mathbf{M}_1$, while for the $FL_1$ dynamics it is anti-parallel to $\mathbf{M}_2$. $M_S$ = 1.2573×10$^6$ A/m [54] is the saturation magnetization of CoFeB based FL, $\mathbf{B}^i_{Eff}$ is the effective magnetic field acting on the $i^{th}$ layer obtained as (superscript *i* is implied, wherever applicable, in all equations below),

$$\mathbf{B}_{Eff} = \mathbf{B}_{Applied} + \mathbf{B}_K + \mathbf{B}_{Dipole} - \mathbf{B}_{Demag} + \mathbf{B}_T \quad (2)$$

where $\mathbf{B}_{Applied}$ is a external bias magnetic field and set to zero vector in this work, $\mathbf{B}_K$ is a uniaxial crystalline anisotropy field obtained as,

$$\mathbf{B}_K = \frac{2}{M_S}\left(K_{Bulk} + \frac{K_{Interface}}{t_{FL}}\right)[0\ \ 0\ \ m_z]^T \quad (3)$$

where $K_{Bulk}$ = 2.245×10$^5$ J-m$^{-3}$ and $K_{Interface}$ = 1.286×10$^{-3}$ J-m$^{-2}$ [54] are the bulk and interface anisotropy constant respectively, $t_{FL}$ is the free-layer thickness in meter, $\mathbf{B}_{Dipole}$ is the dipolar field acting on the $i^{th}$ layer due to the presence of the $j^{th}$ layer and expressed as,

$$\mathbf{B}^i_{Dipole} = \mu_0 M_S \begin{bmatrix} D_{xx} & D_{xy} & D_{xz} \\ D_{yx} & D_{yy} & D_{yz} \\ D_{zx} & D_{zy} & D_{zz} \end{bmatrix} \begin{bmatrix} m^j_x \\ m^j_y \\ m^j_z \end{bmatrix} \quad (4)$$

where **D** is the 3×3 dipolar-tensor and is calculated directly via analytical expressions from Ref. [55, 56]. $\mathbf{B}_{Demag}$ is the self-demagnetizing field of the FL, whose vector is expressed as,

$$\mathbf{B}_{Demag} = \mu_0 M_S \begin{bmatrix} N_x m_x & N_y m_y & N_z m_z \end{bmatrix}^T \quad (5)$$

where $N_x$, $N_y$ and $N_z$ is computed via expressions in Ref. [57]. The thermal fluctuation field which induces magnetic noise in both the FLs is computed as [50, 58],

$$\mathbf{B_T} = \sqrt{\frac{2\alpha K_B T}{(1+\alpha^2)\gamma M_S Vol_{FL} \Delta t}} \begin{bmatrix} R_{0,1}^x & R_{0,1}^y & R_{0,1}^z \end{bmatrix}^T \tag{6}$$

where $K_B$ is the Boltzmann constant, $Vol_{FL}$ is the volume of the FL. Along x, y and z-axis, independent random numbers $R_{0,1}$ have Gaussian distribution of zero mean and unit standard deviation. The thermal fluctuation field is accounted at every time-step. Here, we note that the current induced self-heating effects [59-61] have not been included in this work. For high quality MgO MTJs, which lack metallic pinholes in the tunnel junction either due to the fabrication process or a small cross-section of the MTJ [62], it was experimentally proven [63, 64] that joule heating is not the underlying mechanism for the dielectric breakdown. Instead it is the strong electric field larger than 2 V/nm across MgO insulator [63] that results in the dielectric breakdown. Since, in this work, we ensure that the electric field is always weaker than 1 V/nm and also picked parameters from a high-quality pMTJ [52], these secondary effects are ignored in our analysis. Nevertheless, in case the joule heating dominates or is substantial enough to affect the oscillator characteristics, the characteristics should in general degrade because of the increase in the magnetic noise and the slight degradation in the spin-polarization of the carriers [65]. Next, the MTJ resistance $R_{MTJ}$ calculation accounts for the voltage dependence of the TMR and the dynamic angle $\theta$ between the two precessing FLs as,

$$R_{MTJ} = R_P + \frac{R_{AP0} - R_P}{1 + \frac{V_{MTJ}^2}{V_{Half}^2}} \left( \frac{1 - \cos(\theta)}{2} \right) \tag{7}$$

where $V_{Half} = 0.4$ V, extracted for CoFeB from Ref. [66], is the voltage across the MTJ at which the TMR becomes half of its value at the zero-bias i.e. $TMR_0/2$. $R_P$ is the MTJ resistance when both the magnets are exactly parallel to each other and remains invariant to $V_{MTJ}$ for all practical purpose [67, 68], while $R_{AP0}$ is the MTJ resistance when both the magnets are exactly anti-parallel to each other at the zero-bias. For both FLs, an equal spin polarization P is assumed and computed from the Julliere formula for TMR [69] as $TMR_0 = 2 P^2/(1-P^2)$. The STT efficiency $\eta_{i>j}$ for spin-flux from the layer 'i' to 'j' is computed via Slonczewski expression for MTJ with both sandwiching ferromagnets (FMs) of equal polarization [70],

$$\eta_{1>2} = -\eta_{2>1} = \frac{P/2}{1 + P^2 \cos(\theta)} \quad (8)$$

The expression for $\eta$ with multiple reflections of the spin-flux has been developed for spin-valves [71-73], as noted in Ref. [74] it may not be applicable on MTJs. Next, the DLT with linear dependence on the $V_{MTJ}$ and FLT with the quadratic dependence on the $V_{MTJ}$ are respectively obtained as,

$$a_\parallel^i = \frac{\hbar}{2e} \frac{\eta_{j>i}}{Vol_{FL}^i R_{MTJ}} V_{MTJ} \quad , \quad a_\perp^i = \nu \frac{\hbar}{2e} \frac{\eta_{j>i}}{Vol_{FL}^i R_{MTJ}} V_{MTJ}^2 \quad (9)$$

where $e$ is the electronic charge and $\nu = 2.97/7.82$ V$^{-1}$ is the ratio between the torques extracted from Ref. [67] (for range of values in literature see Table 2.1 and 2.2 of thesis from Kerstin Bernert [75]).

The voltage-dependent MTJ resistance is coded in Verilog-A. The output and transfer characteristics of the high performance NMOS are extracted from the following experimental data from the major foundries presented in IEDM (IEEE International Electron Device Meeting) over the years: 130 nm node bulk-NMOS data from Fig.4 and Fig. 6 of Intel transistors published in Ref. [76], 65 nm node bulk-NMOS data from Fig. 6 and Fig. 7 of IBM, Chartered and Infineon transistors published in Ref. [77], while 14 nm node FinFET (Fin Field Effect Transistor) from Fig. 5 and Fig. 6 of Intel transistors published in Ref. [78]. The channel area for 130 nm and 65 nm is computed as product of channel-length times channel-width, while for 14 nm FinFET its gate length times fin pitch times number of fins (n), where n is an integer not less than the real number obtained by dividing effective channel-width by fin-width (fin-width = 2 fin-height + fin-thickness). The predictive technology files (PTM) available from University of California Berkley (UCB) have been optimized to fit the transistor characteristics. The transistors are modeled via open-source BSIM4 Verilog-A code from Silvaco for bulk-NMOS and from UCB for FinFETs. The circuit is simulated via SPICE solver SMASH 6.5.0 from Dolphin Solutions. The fitting (red line) of NMOS output characteristics via spice simulation against the extracted data (black circles) is shown for instance in Fig. 2(c) for 130 nm node, Fig. 2(d) for 65 nm node and Fig. 2(e) for 14 nm node. For the VCO configuration, the 65 nm node with $V_{DD}$ = 1.1 V, $V_{GS}$ = 1.1 V and channel width to length ratio (W/L) = 21 for an NMOS, while for the CCO configuration $I_0$ = 500 µA is chosen as a default configuration unless specified otherwise.

## III. RESULTS and DISCUSSION

### A. Operation Principle

Figure 1 illustrates the operation principle of the proposed design. For an electron flow along the +z-axis, the transmitted spin-flux oriented parallel to **M₁** acts on FL₂ to attempt to align FL₂ parallel to FL₁, while the reflected spin-flux from FL₂ which is anti-parallel to **M₂**, attempts to align FL₁ anti-parallel to FL₂. In addition to the STT effect, the z-component of the dipolar-field between both FLs attempts to align them in parallel, while the x- and y- components of the dipolar field which participate in the dynamics when the FLs are not aligned along the z-axis attempt to align the FLs in the anti-parallel configuration. Moreover, these two sets of contentions, anisotropy field attempts to drive the FLs along the + or −z-axis depending on if **M** is above or below the x-y plane, respectively, while the demagnetizing field opposes the magnetization vectors in the respective FLs. For reflected flux to effectuate the switching, FL₁ should be weaker than FL₂. This implies that the critical current for switching $FL_1$ as free-layer in the traditional pMTJ STT device should be less than the value that would have been obtained for $FL_2$ as a free-layer. This criterion implies that $FL_1$ should have weaker PMA (perpendicular anisotropy), i.e. it should be thicker, than $FL_2$. For a certain range of values for the physical and electronic constraints, the above effects enable complete and self-sustained successive switching of both the FLs despite of a unidirectional current, as illustrated in the 3D dynamics of the FLs in Fig. 3(a). This is in contrast to a traditional STT device where switching FL into an anti-parallel state w.r.t. reference layer necessitates a current reversal. The set of $FL_1$ and $FL_2$ periodically move across up-up (↑↑), to down-up (↓↑), to down-down (↓↓) to up-down (↑↓) state, where up (↑) and down (↓) refer to magnetization along the + or −z-axis, respectively, which results in a resistance switch between the parallel and the anti-parallel state, as illustrated in Fig. 3(b). When FLs are in an up-up state, the up-spins act on $FL_2$ but the down-spins act on $FL_1$. In this case, the $FL_2$ is receiving spins of the same orientation as its magnetization and hence, $FL_2$ remains unaffected. However, for $FL_1$, if the current is above the critical current required to switch $FL_1$ into an anti-parallel state, after a certain incubation delay, $FL_1$ would start to flip to drive pMTJ towards a down-up state. In this state, $FL_1$ receives the spin-flux oriented along the same direction as $\mathbf{M_1}$ and hence is unaffected. Nevertheless, this transition dynamically starts to change the spin-vector received by $FL_2$ which now tries to follow switching of $FL_1$ to align in parallel with it and attempts to drive the pMTJ into a down-down state. At the same time, the reflected spin-vector also changes because of the instantaneous change in $\mathbf{M_2}$ and attempts to push $FL_1$ to again drive pMTJ into an up-down state. These competing effects in asymmetrically designed FLs ensure that FLs flip each other in stable self-sustained oscillations. We would like to emphasize that this explanation is somewhat oversimplified to enable conceptual understanding of the operation. As can be gauged from the dynamics in Fig. 3(a), the transient magnetization states would rarely be in a clear set of up and down states. Mostly they would have a significant x and y component, thus empowering the contentious effects of the dipolar-coupling to play a significant role in the dynamics. Therefore, as evident in Fig. 3(b), one FL may start to switch before another FL is fully switched. The FL dynamics is, on the other hand, such that although **M₁** and **M₂** mostly do not simultaneously exactly align with the z-axis, they do achieve exact parallel and anti-parallel states. This enables the design to achieve a full resistance swing from $R_P$ to $R_{AP}$ and vice-versa. This design attains theoretical maximum swing, with a constant current input, that can be obtained via STT oscillation, which to the best of our knowledge is not possible even in the STNOs with the in-plane FL with a perpendicular polarizer (pi-Fixed) and the perpendicular FL with an in-plane reference layer (ip-Fixed) [20, 41, 79, 80], indicating a novel concept of obtaining oscillations.

These oscillations in resistance electronically translate to the voltage oscillations across a MTJ as shown in Fig. 3(c). The change in $R_{AP}$ and $R_{MTJ}$ as the voltage across the MTJ changes is taken into account as discussed in the methodology section via eq. (7). FFT of the output $V_{MTJ}$ is shown in Fig. 3(d) and 3(e) in dBm (to show output power) and mV units respectively. Figure 3(d) also shows the first two overtones at the multiples of fundamental tone $f_0$. The frequency at which the largest output voltage $V_{Pk}$, i.e. the largest output power, is obtained corresponds to $f_0$. A 3 dBm line below the peak value (dashed black line) subsequently gives the half-power frequencies $f_1$ and $f_2$ on the either side of $f_0$. The FFT plot in mV units is manually fitted with the Lorentzian function [25, 81] $V_{Pk}/(1+((f-f_0)/\Gamma)^2)$ (blue line in Fig. 3(e)), where $\Gamma$ is a fitting parameter. The intercept of the function with the −3 dBm line gives $f_1$ and $f_2$, linewidth $f_2-f_1$, and quality-factor (Q) which equals $f_0/(f_1-f_2)$. In Fig. 3(c), an abnormal observation in this design is the strong amplitude-fluctuations for small $V_{MTJ}$ i.e. when the FLs are or nearly parallel to each other. These fluctuations occur because of the strong dipolar coupling between the FLs which tends to lock them together.

This assisted with a net FLT makes one FL precess around another while DLT tries to align them via damping effect, governed by the damping factor α. Increasing the α, weakening the dipolar field or weakening the FLT reduces this noise, but the frequency of the oscillator also reduces. Furthermore, for a very high frequency at the large currents, i.e. for very fast switching, the overlap between the dynamics of $FL_1$ and $FL_2$ is such that the transient angle between the FLs do not reach full 0 and 180 degree, rather the window starts to diminish which degrades the device performance as shown later. Hence, further research into reducing these amplitude-fluctuations (noise), without compromising with the frequency, would significantly increase the output power $P_O$ of the SW-STO and thus would be an important future direction.

**B. Magnetic Noise**

Next, we investigate the effect of the thermal fluctuation field that induces a random phase to the magnetization dynamics [82] of both the FLs. Thermal fluctuation field results in a Gaussian noise and modeled by eq. (6). It strongly depends on the volume of the magnet $Vol_{FL}$. Figure 4(a) shows that despite of magnetic noise both the FLs of the SW-STO still switch and oscillate between $m_z = -1$ and $+1$. Figure 4(d), 4(e) and 4(f) respectively illustrate the 3D magnetization dynamics and FFT of the $V_{MTJ}$ in dBm and mV for an ip-MTJ based STNO of 60nm×40nm×5nm(1.8nm) dimensions, with a fixed in-plane pinned layer in place of the $FL_1$ in absence of magnetic noise, while the same is respectively illustrated in Fig. 4(g), 4(h) and 4(i) with magnetic noise. This device can also be operated without any external bias field and generate oscillations in the microwave range. Figure 4 illustrates that although a magnetic noise adversely affects the oscillator performance, the SW-STO is much more immune to the thermal field than the precessional orbit-based STNO. To show this, the effect of magnetic noise is compared on SW-STO and a precessional orbit-based STNO (see trajectory in Fig. 4(d, g)) made of MTJ with same RA-product, FL of similar volume and delivering same output power (−22.5 dBm or 23.7 mV, compare peak of Fig. 3(d) with 4(e) or 3(e) with 4(f)) in the absence of magnetic noise. For SW-STO, the peak output power $P_O$ drops by −10.64 dBm in Fig. 4(b) w.r.t. peak $P_O$ in Fig. 3(d) without the thermal field. The $V_{Pk}$ goes down to 6.8 mV and linewidth increases from 1.67 MHz to 28.16 MHz i.e. Q degrades from 204 to 12.43. This degradation in oscillator characteristics, however, is quite modest to what is observed for the precessional orbit based STNOs. Magnetic noise strongly impairs the orbit as shown in Fig. 4(g) which degrades the peak $P_O$ by −25.38 dBm (compare peaks in Fig. 4(e) with 4(h)), the peak voltage $V_{Pk}$ goes down to 1.3 mV and the linewidth increases from 0.0874 MHz to 335.7 MHz i.e. Q degrades from 58760 to 15.33 for the STNO. The power gets distributed into the adjacent frequencies which broadens the linewidth and degrades the Q factor. Also compare the output power and the voltage in the presence of the magnetic noise between SW-STO and STNO in Fig. 4(b, c) with Fig. 4(h, i) respectively to observe the relative robustness of SW-STO w.r.t. traditional precessional orbit based STNO. It is found that the SW-STO achieves relatively more thermal stability because of the complete switching into the highly stable states (z = ±1) along the z-axis where both of the layers spend some time (see flat portion of the waveforms in Fig. 3(b) and 4(a)) before continuing with their transience to toggle again. During these rest periods the anisotropy fields are strong, while the in-plane demagnetizing field that would have pushed away the alignment of the magnetization from the z-axis is nearly zero. For a brief period when the two FLs are in a fully switched AP-state, even the spin-torque due to the current is practically zero. In contrast to STNOs, the attainment of the stable state also allows the device to attain a stable resistance value for the MTJ for a certain duration and that too when the thermal fluctuation would have weakest effect on the dynamics of the oscillator. For STNOs, since the magnetic noise impairs the orbit of the magnetization vectors, it results in a stronger degeneration of the precise dynamic angle between the FLs thereby affecting the resistance and subsequently the $V_{MTJ}$ oscillations. Since, magnetic noise affects both the amplitude-noise and the phase-noise (quality factor) of the spintronic oscillators, the transient characteristics of a STNO become more chaotic than for a SW-STO.

Since, in this work, switching based design is investigated for only a free-running case, for a rightful evaluation, we compare only with a free-running STNO. Injection locking or synchronization of the magnetization dynamics in the STNOs [24, 25] and the phase-locking in the ring-oscillators [42] dramatically improves their Q, and may have similar benign effect on the proposed SW-STO. Furthermore, since when FLs are nearly locked parallel with each other and one FL starts to precess around the other FL resulting in the rapid oscillations (c.f. Fig. 3(b)), some technique to suppress or rapidly damp these oscillations should greatly enhance the quality-factor of the SW-STO. Investigating these techniques, however, is beyond the scope of the current work and could be a possible future direction to further the investigation of this design. Therefore, for the precessional orbit-based designs (free-running case) scaling down the MTJ dimensions has much worse and stronger effect than on a switching based design which thus withstands better chances of down-scaling of the magnets.

As noted earlier in operation mechanism, there is a physical constraint in pairing of the thickness of the FLs which determines the operability of the design. As a result, Fig. 5 appraises the effect of the FL thickness on the device operation to show that the design can be optimized to meet the frequency requirements over a range of specifications. To reiterate, for a device to work as per our chosen current direction, $FL_1$ should be weaker (in this case thicker and hence closer to the critical thickness at which magnet would become in-plane) than $FL_2$, so that $FL_1$ can switch from the reflected spin-flux. As observed from Fig. 5, a smaller difference in the thickness of the two FLs increases the frequency. This, however, reduces the stability of the oscillations such that they die out as the two thicknesses approach each other. As a result, for a smaller difference in the thickness, although frequency increases, the current controlled tunable range degrades.

**C. CCO Configuration**

Figure 6 presents the oscillator characteristics for operation in a CCO configuration. To focus only on the MTJ dynamics, we do not account for the power, area and the design of the current sources/mirrors which would drive these oscillators, in sync with the STNO literature where it is dealt separately [8, 83]. Integration with the driving transistor(s) will be addressed in the next sub-section. In Fig. 6(a), a wide tuning range from 71 MHz at 195 µA to 965 MHz at 1100 µA (red stars) i.e. a tuning ratio of 13.6 is presented. More detailed comparison of the entire performance metrics is given in Table I. It can be found that it is much larger than the values reported for STNOs for current control [20, 41, 79, 80] and the ring-oscillators [42]. The output $V_{MTJ}$ (blue circles) roll-down from −26.24 dBm at 172.5 MHz to −47.7 dBm at 965 MHz. As the current increases, the voltage drop across MTJ also increases. This reduces $R_{AP}$, the maximum resistance of the MTJ for a given operating bias for the anti-parallel state. As a result, the maximum attainable swing for the resistance oscillations decreases as the voltage drop across the MTJ increases. Therefore, despite of the increase in the current, the $V_{MTJ}$ swing decreases. On the other hand, if there had been no such dependence of the TMR on the voltage bias, the $V_{MTJ}$ swing would have increased. In fact this is indeed observed for small enough currents in Fig. 6(a) (blue stars) where increase in the current over powers the reduction in $R_{AP}$ to result in an increasing trend for the $P_O$ initially. The effectiveness of the proposed design, nevertheless, becomes more evident on computing the power consumed (red stars) and the power efficiency (blue circles), illustrated in Fig. 6(b). With an increasing current, as would be expected, the $P_{IN}$ increases and the efficiency decreases. Efficiency of 0.12%, i.e. nearly −3 on the logarithmic scale, for STNO [41] corresponds to −42 dBm of output, while for the proposed design, at this efficiency, the output is −33.5 dBm.

Next, Fig. 6(c) demonstrates that due to the increased noise and the amplitude fluctuations at the large currents, and hence at the higher frequencies, the linewidth (red stars) increases, thereby degrading the Q (blue circles). The linewidth is however also observed to increase at very small frequencies and resulting in small Q. The trend shows weak minima for the linewidth trend and strong maxima for Q-factor in lower half frequency regime, i.e. the best performance is observed in this region and degrades as the current is either reduced or increased. These trends can be understood as follows. As mentioned above in the operation mechanism, at the larger frequencies at the larger currents corresponding to the larger control voltages, the dynamic angle θ between FLs may not reach full 0 and 180 degree and starts to diminish. Alternatively, it can be understood as an increase in the overlap between the transient characteristic of the $m_z$ of the two FLs or the asymmetric phase shift in the transients. In other words, the larger currents trigger faster switching of the two FLs which are affected asymmetrically because of their asymmetric design (different thickness for the same ferromagnetic parameters). This not only slightly reduces the $V_O$ but also increases the importance of the traditional precessional trajectories, and concurrently increases the fraction of the time per cycle spent by one FL to process around another when they are nearly parallel. The increased weightage or the importance of the precession thus increases the importance or the influence of the magnetic noise on the oscillations, which results in the increased linewidth and the reduced quality factor. On the other hand, as the current decreases, the MTJ current reduces and eventually approaches the value of the critical current required for the switching of each FL. This has following two key implications. First, as expected, the incubation time for the onset of the oscillations, i.e. triggering the oscillator, increases (not shown) due to weak current. The onset time becomes strongly stochastic due to the magnetic noise which eventually triggers the oscillator. This would have important implications on the write time of a STT memory, but is not a concern for a tunable oscillator because this falls within the wakeup or startup time of the circuit or a chip. The edges of the clock serve as a reference point for measuring time and thus anything before the clock has stabilized becomes immaterial. Second, the incubation-delay from period-to-period in the oscillations becomes more evident, and in fact determines the lower bound of the obtainable frequency. Period-to-period variations happen because the switching is now slow enough to have the FLs in the perfect parallel and anti-parallel states with both **$M_1$** and **$M_2$** aligned along the z-axis long enough, rendering the torques to a near zero such that the following effects play a significant role. The role of the dipolar-field becomes important to unlock from these conditions, especially when the FLs are in an anti-parallel configuration. However, more importantly, the dependence on the random thermal-field (magnetic noise) increases to unlock the FLs from these alignments. This becomes the key cause of the period-to-period variations from the oscillations between the $R_P$ and the $R_{AP}$ states, since the random thermal-fields introduce random angle-offsets in $\theta$ making one switching cycle faster than the other. Therefore, the linewidth increases and the quality factor decreases for very weak currents. The difference in strength of inflection on the trend for linewidth and quality factor (Q) emerges because later also depends on $f_0$. This $f_0$ which comes in the numerator of the expression for computing Q is small for a small driving current. Hence, the small change in the denominator i.e. linewidth is more strongly reflected in Q. This results in strong maxima for Q for weak minima in linewidth in lower half of the operational window of the oscillator. The points of inflection, however, do not align due to dependence of Q on $f_0$. Nevertheless, the observed Q is in the range of 4.2 to 21, which is comparable to the Q-factor for the free-running STNOs [41]. This again affirms the usefulness of the proposed design.

Next, since we have specifically emphasized on the large tunability ratio of the proposed device and here in the CCO configuration shown a ratio of 13.6 which is much larger than other solutions (see Table I for comparison), we explain here the importance of this performance metric. An important issue with the extant solutions is their small tunability ratio, for both semiconductor and spin-based solutions [41, 42, 84]. For the multi-band operation, directly driven by the tunable oscillator (CCO/VCO/DCO), the advantage of a large tunability ratio is implicit. However, in many applications tunable oscillator is integrated within the PLL which then generates the oscillations of the desired frequency, like for clocks in VLSI chips and local oscillator signal for transceivers. Here, despite of a small tunability ratio, even as large as only an octave, of the oscillator it can be propounded that a frequency down-counter of the larger values can be used to achieve even smaller frequencies and thus artificially expand the tunability range with an insignificant increase in the area and cost. It thus appears that there is no need of a large tunability ratio in an integrated tunable oscillator. However, actually there is indeed a need of a large tunability ratio, but not implicit in these systems. It can be understood in detail as follows. In PLL, a divider (frequency down-counter) is implemented in the feedback loop. This divider scales down the frequency by a natural number N programmed into the divider. The output of this divider is thus used as an input to the phase-detector whose other input would be coming from a stable crystal oscillator of generally much lower frequency. Designing a larger N-value divider does not incur much additional area or cost penalty as well. Similarly, even at the output of PLL another frequency down-counter (divide by M-value) can be used, with insignificant increase in the cost and area, which finally gives the desired clock frequency. Hence, an octave band oscillator seems to suffice, provided it can generate large enough frequency that can be down-scaled. This oscillator output can be thus be divided by N and M respectively to generate input for phase-detector and the clock output. Using a large frequency output of the oscillator (for instance with an octave tunability) just requires a larger value of N and M, which insignificantly increases the cost and area. Unfortunately, not only for the proposed SW-STO (as we show later in results section) but even in general the power consumed by the oscillator increases significantly at the large frequencies. For instance, for a ring-oscillator the dynamic power is directly proportional to the frequency, and the quality-factor degrades (or phase noise increases) tremendously at larger frequencies. Hence, designing a tunable oscillator at much higher frequency than the desired clock frequency strongly penalizes the performance with respect to (w.r.t.) to both the power and the phase noise, and thus the design of the PLL would suffer. Hence, operating the oscillator as close as possible to the clock frequency is a standard design technique. Hence, even for PLL driven multi-band operation, a large tunability ratio is a very much desired characteristic of a tunable oscillator. A decade-wide tunable oscillator can therefore operate with a smaller value of N and M by generating frequency closer to the desired clock output over a wider frequency spectrum and can thus significantly reduce the power consumed in the PLL.

### D. VCO Configuration

### D.1 Driven by NMOS

From the practical point of view, for the CCO configuration, the MTJ needs to be driven by a tunable constant current source. Designing a current source in turn would need a few transistors. MTJ is fabricated among higher metal layers as part of BEOL (back-end of line) process while these driving transistors are fabricated during FEOL (front-end of line) process. Therefore, topologically, the transistors are laid out under the MTJ. Hence, the larger of the two areas (MTJ or the driving circuit) gives the actual area occupied by the cell. This indicates that only comparing the MTJ area with that of the ring-oscillator to proclaim area advantage for former is not correct. This is an important design consideration often overlooked in a spin-oscillator literature. Furthermore, it becomes more difficult to design stable widely tunable current sources/mirrors, especially at smaller CMOS nodes [8, 83], than designing VCO/CCO with the wide tuning range. Therefore, we consider the case of driving a MTJ with a single gate-controlled NMOS in this section. Figure 7 illustrates the effect of the transition from a CCO to a gate-controlled VCO configuration. The supply voltage $V_{DD}$ and the gate voltage are fixed whereas the node connecting MTJ with the drain end of the NMOS is floating. The voltage at this node is now the output voltage $V_O$ that would be routed to the subsequent stages driven by the oscillator. Consequently, in a VCO configuration, we compute $P_O$ corresponding to the $V_O$ instead of the $V_{MTJ}$. It is understood that the oscillatory or the AC component of $V_O$ is just inverted w.r.t. AC component of the $V_{MTJ}$, and they are different only w.r.t. a DC operating point. As a result, to have a proper comparison of the voltage oscillations between a VCO and a CCO configuration, $V_O$ in Fig. 7, for a constant current source, is inverted by subtracting $V_{MTJ}$ from 1.1 V which is the $V_{DD}$ used for the VCO configuration. Figure 7(a, b) show the $R_{MTJ}$ and $V_O$ for the CCO configuration for comparison with the VCO configuration operating at 350 MHz. To attain this, the CCO configuration is operated at the current equal to the average current observed for the VCO configuration which is 450 µA in this case. As the resistance of the MTJ oscillates, it changes the quiescent or the DC operating point of the drain end of the NMOS. Therefore, it can be found that not only the voltage across the MTJ oscillates but the current through the cell also oscillates, as shown in Fig. 7(c), in anti-phase with $R_{MTJ}$ (Fig. 7(d)) i.e. characteristics of the $I_{MTJ}$ and $R_{MTJ}$ are inverted w.r.t. each other; however, $V_O$ (Fig. 7(e)) oscillates in-phase with the $R_{MTJ}$. The anti-phase oscillations diminish the output voltage swing. Peak-to-peak voltage swing $\Delta V$ can be approximately given by,

$$|\Delta V| \sim |I_{AVG} \cdot \Delta R - R_{AVG} \cdot \Delta I| \qquad (10)$$

where $I_{AVG}$ is the average of the oscillating current $I_{MTJ}$ through the cell, $R_{AVG}$ is the average resistance of the oscillating $R_{MTJ}$, and $\Delta I$ and $\Delta R$ are the respective peak-to-peak swings in the current and the resistance oscillations. Equation (10) is mathematically implicit from taking partial derivative of $V_{DS} = V_{DD} - I_{MTJ} \cdot R_{MTJ}$ with proper phase for all three variables. It can be observed that $\Delta V$ is not equal to twice of $V_O$. The magnitude of $\Delta V$ includes the amplitude-noise while $V_O$ is the amplitude of the swing over the noise floor. Therefore, $V_O$ cannot be extracted manually from the noisy data, and FFT is performed on the obtained data to estimate the $V_O$. Since, $V_O$ is the actual usable signal, after subtracting for the effect of the noise, corresponding usable peak-to-peak value i.e. $2 \cdot V_O$ is smaller than $\Delta V$. Since $\Delta V$ can be calculated even manually from simple arithmetic operation on the transient plots of $I_{MTJ}$ and $R_{MTJ}$, and thus more intuitive, it serves as a simple tool to conceptually understand the VCO characteristics, presented in subsequent sub-sections. Equation (10) directly implies that the magnitude of the $\Delta V$ (which includes noise) for the proposed oscillator, on being driven in a VCO configuration decreases w.r.t. a CCO configuration because $\Delta I$ is zero for the later. Consequently, $V_O$ extracted from the noisy data also reduces for the VCO configuration, thereby degrading the output power $P_O$. Although explicit analysis has not been done for traditional STNO driven by a NMOS like for the proposed design, given the same concept applicable at the drain node for STNO in series with NMOS, we expect similar drop in $P_O$ for the STNO case as well in the same configuration.

Another conceptual approach that should assist in understanding the results in subsequent sub-sections is the load-line analysis (Fig. 8). Load-line analysis is a standard approach [85] to graphically obtain the DC bias or the quiescent point of the system shown in Fig. 2(a), i.e. it gives the DC current through the NMOS and the DC voltage at the drain terminal. In Fig. 8, for simplicity, $R_{MTJ}$ is treated as a constant i.e. a trivial resistor. As shown in Fig. 8(a), for the gate-control or $V_{GS}$ sweep with a fixed $V_{DD}$, the output characteristics of the NMOS (solid line) i.e. current $I_{DS}$ vs. drain-to-source voltage $V_{DS}$ characteristic for an increasing gate-to-source voltage $V_{GS}$ is intersected with a single dashed line of the output characteristic of a resistor which intercepts the y-axis at $V_{DD}/R_{MTJ}$ and x-axis at $V_{DD}$, giving a slope of $-1/R_{MTJ}$. The intersections of these two output characteristics are marked via red dots and give the quiescent point or the operating DC conditions. It can be found that as the $V_{GS}$ increases, the DC voltage at the drain terminal reduce while the MTJ or the drain current increases. In contrast, for the $V_{DD}$ sweep with a fixed gate voltage, shown in Fig. 8(b), load-line analysis shows a single line (actually superimposed, with larger $V_{DD}$ line overshadowing the smaller $V_{DD}$ trend line) for the NMOS output characteristic, and multiple parallel lines for $R_{MTJ}$ of $-1/R_{MTJ}$ slope, one each for an increasing supply voltage $V_{DD}$. This results in the increase of both the DC voltage at the drain terminal and the drain current when the $V_{DD}$ increases. These concepts would be used in section D.2. Next, when both $V_{DD}$ and $V_{GS}$ are fixed in Fig. 8(c) but the resistance of the MTJ is swept, a single line for the NMOS output characteristic is intersected via respective $R_{MTJ}$ output characteristic line with a corresponding slope. This results in the decrease of both the DC voltage at the drain terminal and the drain current when $R_{MTJ}$ increases. This concept would be useful in understanding section D.3.

### D.2. $V_{GS}$ and $V_{DD}$ Control

Figure 9-11 show the control of the oscillator via tuning the gate-voltage $V_{GS}$ and the node voltage $V_{DD}$. Although both methods show nearly a same frequency tuning range in Fig. 9 and same range of linewidth and quality factor in Fig. 11, of the two methods, superior characteristics like a larger $V_O$ (Fig. 9) and better power efficiency $P_O/P_{IN}$ (Fig. 10) are exhibited for the $V_{GS}$ control of the NMOS. Henceforth, the detailed mechanism behind the observations, their impact on the oscillator performance and the trade-offs for controlling via $V_{GS}$ or $V_{DD}$ are expounded in this sub-section.

***Impact on Frequency***: For both the gate and the $V_{DD}$ control, the current increases with the voltage which subsequently increases the oscillation frequency (see red stars in Fig. 9). However, it is also observed that the frequency linearly increases on tuning the $V_{DD}$, while it saturates for a large $V_{GS}$. This can be understood with the load-line analysis described earlier. In Fig. 8(a), for the gate-control, the quiescent point is in the triode region. As $V_{GS}$ increases, the operational $V_{DS}$ decreases and the $I_{DS}$ increases and both eventually saturate. Conversely, in Fig. 8(b), for the $V_{DD}$ control, if the quiescent point (red dots) enter the saturation region of the transistor, the current has almost no change (slightly increase due to the non-saturating behavior at smaller nodes due to the short-channel effects), but $V_{DS}$ would continue to increase. If a quiescent point is in the triode region, both $I_{DS}$ and $V_{DS}$ would increase continuously. Due to the large $R_{MTJ}$ and a relatively large current requirement to switch the FLs, the voltage drop across the MTJ forces the transistor to operate in the triode region in this work. Following this operating or the DC bias of the system with the NMOS in the triode region, and the current $I_{DS}$ and the drain voltage $V_{DS}$ following the explanation above, the frequency (c.f. Fig. 9; red stars) continuously increases for the $V_{DD}$ control while it saturates for the gate-control of the oscillator.

Thereafter, it can be observed that in a VCO configuration the lowest frequency operable is lower than that in a CCO configuration (see Table I for explicit numbers). At the lowest frequency in the VCO configuration, the average drain current $I_{AVG}$ is lower than the value at which the oscillator works in the CCO configuration. The working of oscillator at such low current can be attributed to the current $I_{DS}$ oscillations in the VCO configuration. During the oscillations, the instantaneous current is much larger than the lowest operating current in the CCO configuration for a substantial fraction of the time-period. This current drive is observed to be still sufficient to trigger the oscillator. At the lowest current drive in the $I_{MTJ}$ oscillation, when the MTJ is in the anti-parallel state, the switching dynamics to transition the MTJ into a parallel state and thus increase the current drive is sustained mainly via dipolar field.

*Impact on Output Power*: The downside of the current oscillations in the VCO configuration is to degrade the $V_O$, as discussed earlier and illustrated in Fig. 7. $V_O$ (blue circles in Fig. 9), however, not only degrades but it also exhibits more complex behavior which can be understood via eq. (10). Succinctly, the increasing trend of $V_O$ is observed when the increase in the current outdoes the fall in the MTJ resistance due to latters dependence on the TMR, whereas a decreasing trend of the $P_O$ is observed just like for a CCO configuration when the increase in the current is insufficient to compensate for the decrease in the $R_{MTJ}$. To obtain the detailed insights, the detailed mechanism is investigated as follows. The current increases as the $V_{GS}$ or the $V_{DD}$ increases, thus $I_{AVG}$ increase in both cases. Because of the increase in $I_{AVG}$, the dc bias across the MTJ also increases (Fig. 8). The rate at which this dc bias changes and the extent to which it can change is however different for the two type of controls. For the $V_{GS}$ control, an increase in $V_{GS}$ results in a downward shift in the drain bias which automatically translates to a larger MTJ bias because of a fixed $V_{DD}$ which in turn is set at its maximum value permissible for the node. For the $V_{DD}$ control, an increase in the $V_{DD}$ results in an upward shift in the drain bias. However, this change in the drain bias is not as swift as the change in $V_{DD}$ which thus allows the MTJ bias to increase ($V_{DD}$ = drain dc bias + MTJ dc bias voltage). This difference in the dynamics of the operating voltages of the drain terminal and the MTJ result in very different trends for $V_O$. The upper bound due to a smaller $V_{DD}$ in the case of a $V_{DD}$ control, despite of an increasing MTJ dc bias, results in a nearly constant $V_O$ with all values within a small window of only 1.5 dBm with a base value of −44dBm. The points seem to be randomly distributed in this narrow range owing to the finer balance among dc bias, MTJ dynamics and TMR roll-off with the MTJ voltage. Averaging over 16 runs for each of the case of $V_{DD}$ control, however, shows a trend with a weak maximum in the center as shown via dashed line in Fig. 9(b). The subsequent finer discussion on the effect of $V_{DD}$ control on $V_O$ therefore implicitly refers to this average behavior. Next, because of the dependence of TMR on $V_{MTJ}$, the $R_{AP}$, $\Delta R$ and $R_{AVG}$ reduce as the current increase in both cases. Since the quiescent point of the drain terminal acts as an additional constraint for regulating the potential drop across the MTJ, it results in constraining of $\Delta R$ and $R_{AVG}$. Consequently, for the $V_{DD}$ control, the $\Delta R \cdot I_{AVG}$ firstly increases and then decreases whereas for the gate-control it always decreases and starts to saturate. The non-linear dependence of the current on the respective voltage control and the non-linear relation between the $R_{MTJ}$ and the $V_{MTJ}$ results in firstly an increasing and then decreasing trend for $\Delta I$ (current oscillations) for the $V_{DD}$ control, whereas it is mainly an increasing trend for the gate-control. With the consideration of $R_{AVG}$ and $\Delta I$ together, for both the $V_{DD}$ and the gate-control, $R_{AVG} \cdot \Delta I$ continuously decreases with an increasing control voltage. Therefore, the operating dc bias, $R_{AVG}$, $\Delta I$, $\Delta R$ and $I_{AVG}$ collude to result in the change in $\Delta V$ with same dependence on the $V_{GS}$ and the $V_{DD}$ control as the trends observed for the $V_O$ or $P_O$ in Fig. 9. For the results of the $V_O$ data, it can also be observed that the gate-control enables a larger $P_O$ again because of the full-scale $V_{DD}$ which equals to the node supply voltage, but a lower $V_{DD}$ because the $V_{DD}$-control for the oscillator inherently limits the $V_O$ and the $P_O$. Conspicuously, irrespective of the control, one shortcoming of this design, compared to a ring-oscillator (see Table I) used for the ASIC clocks, is still much lower output power, indicating that an additional amplifier driven by a $V_O$ signal may still be needed and consume additional area and energy. The design of this amplifier, however, is beyond the purview of the current work and hence the configurations in Table I for each column are prudently specified to bring out this difference explicitly to notice.

*Impact on consumed power and power-efficiency*: Next, the power consumed $P_{IN}$ (red stars) in Fig. 10, as expected, follows the same trend as the current or the frequency for the respective type of voltage controls. A larger $P_O$ for the gate-control, with $P_{IN}$ in the same range as for the $V_{DD}$-control, evidently results in a larger power-efficiency $P_O/P_{IN}$ (blue circles) for the gate-control. Furthermore, for the gate-control, the cell does not load the preceding driving stage or the circuit which generates the control voltage because of the large (theoretically infinite) input impedance of the gate terminal of the transistor. For the $V_{DD}$-control, the drain-to-source impedance is much smaller w.r.t. the gate input impedance and thus loads the driving circuit. Therefore, for practical designs, the gate-control of the oscillator is suggested instead of the $V_{DD}$-control method of the ring-oscillators.

*Impact on linewidth and quality-factor*: In the VCO configuration, there are three key factors which result in noise and thus affect the linewidth. First is the amplitude-noise generated when $R_{MTJ}$ tends to $R_P$ (c.f., Fig. 3(c)) i.e. when the two FLs become nearly parallel, one FL starts to strongly precess around another, resulting in the rapid oscillations. Second is the addition of the magnetic noise which has Gaussian random distribution. Third, is the fluctuation in the operating bias of the drain terminal which further degrades the oscillator characteristics (Fig. 7). Therefore, in Fig. 11, the linewidth increases because of the increasing noise with the increasing voltage for both $V_{DD}$ and $V_{GS}$ control. For the first data point of $V_{DD}$ control the current and corresponding frequency are small enough to mark heavy dependence on the magnetic noise which thus increases the linewidth. The physics here is same as observed and explained for very small currents in the CCO configuration. Lastly, the Q factor, computed as $f_0/(f_1-f_2)$ i.e. the ratio of the fundamental tone to the linewidth, follows the simultaneous effect on the fundamental frequency and the linewidth for both a CCO and a VCO configurations. Hence, at both ends of the current strength, the linewidth is larger and the quality factor is smaller than its adjacent points. The Q is in the range of 3.7 to 13.3, which is slightly lower than the values obtained in a CCO configuration for the SW-STO and for the free-running STNOs [41]. The VCO frequency is now in the range of 47.6 MHz to 276.3 MHz, which limits its applicability to the ASIC clocks, with the tunability ratio of 5.8. This ratio is still much greater than both for the STNOs and the ring-oscillators (see Table I). These results show that even in the VCO configuration with a single NMOS, the proposed design can contest with the effectiveness of the other oscillator designs.

### D.3. MTJ Area

Figure 12 next appraises the effect of scaling the MTJ cross-section for fixed drive strength of the NMOS. As the area of the MTJ is scaled, it proportionately scales the $R_{MTJ}$ because of a constant RA-product. In Fig. 12(a), the oscillator frequency goes down with the increasing area because the current density scale-down almost quadratically with the MTJ dimension. Deviation from the quadratic trend happens because of the changes in the DC bias conditions (see red dots in the load-line analysis illustrated in Fig. 8(c)), due to the change in the $R_{MTJ}$. As observed in Fig. 8(c), as the $R_{MTJ}$ scales-down for the fixed operating condition of the NMOS, the drain current and the DC voltage of the drain terminal increases while the DC bias across the MTJ i.e. approximately the average value of the $V_{MTJ}$ reduces. This increase in the current, however, cannot over-ride the decrease in the current density due to the quadratic effect of the scaling-up of the MTJ dimension. Hence, the frequency (red stars) scales-down but the trend is weaker than the trend that would have been obtained for pure quadratic effect of area scaling. Fixed RA product also results in scaling down of both $\Delta R$ and $R_{AVG}$ which is again weaker because of the smaller DC value of the $V_{MTJ}$ caused by the shift in the bias conditions. The shifts in the bias conditions also allow for much more steep increase in $I_{AVG}$ than in $\Delta I$. This is further assisted positively by an increased thermal stability of the MTJ because the magnetic noise is directly proportional to the square-root of the volume of the magnet. Because of these three reasons, the output power $P_O$ (blue circles) increases with the increasing cross-section but at the cost of the frequency. Increase in the current also results in the increase in $P_{IN}$ (red stars) as shown in Fig. 12(b), but the improvement in $P_O$ still results in a higher power-efficiency (blue circles) for the larger MTJ cross-sections. Furthermore, as expected from the effect of the volume of the magnet on the magnetic noise, the linewidth (red stars) improves i.e. linewidth decreases as the MTJ cross-section increases in Fig. 12(c). The quality factor, $f_0/(f_1−f_2)$, follows the combined trend of frequency and linewidth. Therefore, it shows local maxima in the center and tapers at the two ends, at the lower end because of large linewidth and at upper end due to small operating frequency. For a 50% increase in the dimension from 50 nm to 75 nm or nearly twice the increase in the area of the MTJ, the linewidth improves from 24 to 14 MHz i.e. by 41.7 %, while the Q factor is nearly 12 for both cases. This shows that if for a given frequency the design permits the use of a larger MTJ, it could be extremely advantageous in terms of the output power, power-efficiency and the linewidth. To examine this viability, we next appraise the scaling of the drive strength of the NMOS across three CMOS nodes and understand how they provide the margin for using larger MTJs despite of scaling-down of the CMOS nodes.

### D.4. NMOS W/L Scaling

In Fig. 13, the effect of scaling channel width of NMOS to its length, i.e. W/L ratio, for 14 nm (red stars), 65 nm (blue circles) and 130 nm (magenta squares) CMOS node is studied, integrated with a MTJ of dimensions 50nm×50nm×1.6nm(1.2nm) i.e. a cross-sectional area of 0.0025 μm$^2$. As expected the average current $I_{AVG}$ for all three nodes increases with the increasing W/L in Fig. 13(a), and almost saturates because of the integration of the NMOS with the MTJ resistor with determines the DC bias of the system. The channel area is computed based on the approach introduced in the methodology section and illustrated for reference in Fig. 13(b). For 14 nm node, the channel-area is from 0.00168 μm$^2$ to 0.0468 μm$^2$, for 65 nm node it is from 0.009245 μm$^2$ to 0.194 μm$^2$, and for 130 nm node from 0.0245 µm² to 0.5145 µm² for W/L ratio from 5 to 105, respectively. From Fig. 13(a, c), it is found that for 14 nm node, $I_{AVG}$ is from 200 µA at 86.32 MHz to 284 µA at 160.7 MHz for W/L from 10 to 105, for 65 nm node it is from 237.4 µA at 119.2 MHz to 507.8 µA at 357.3 MHz for W/L from 7 to 105, and for 130 nm node from 338.1 µA at 202.2 MHz to 630.2 µA at 486.6 MHz for W/L from 5 to 105, respectively. The node supply-voltage $V_{DD}$ is 0.7 V, 1.1 V and 1.3 V, respectively for the three nodes. Hence, it can be observed from the data that MTJ is always smaller than the channel area of the NMOS. Furthermore, the channel-area of the NMOS is smaller than the actual area of the NMOS which includes the area for the drain and the source region. This shows that firstly the design area would be dominated by the transistor instead of the MTJ and thus should serve as the rightful parameter for comparing area with other designs like the ring oscillators. This fact has often being ignored in the STNO literature. The power and the area of the driving unit have been ignored as well while comparing the designs. Secondly, within the periphery of the folded (because of a large W/L) NMOS, a larger MTJ can be used without any additional penalty on the design area. Therefore, this W/L analysis for the NMOS over three CMOS nodes shows the viability of using larger MTJs to boost the performance of the oscillator. These results also show that the MTJ technology does not need to directly scale-down or at least at the same pace with the CMOS technology for the proposed oscillator. Next, the comparison across the nodes in Fig. 13(d) shows that using larger nodes for the oscillator design would be beneficial as the nominal supply voltage of the node also scales down with it thereby reducing the drive capability of the transistors. On the other hand, the data also implies that scaling-down the CMOS node limits the operational range of the oscillator and for a given current (iso-frequency) degrades the output and consumes more channel area, for instance compare approximately between the last point of 14 nm and the second point of 65 nm trends.

**D.5. Digital Control: DCO Configuration**

From the frequency range shown in Fig. 13(c), a tunability ratio cannot be computed because channel width is not normally electronically tunable in a VCO or a CCO design. This constraint can be circumvented and channel widths can be electronically controlled by re-configuring a VCO into a DCO. Since the W/L analysis shows that a large channel width for the bulk-NMOS or a large numbers of fins for the FinFETs are required to operate this design, it provides the ability to extend the VCO based analog design to a coarse DCO based digital design, as illustrated in Fig. 14(a). For an N-bit digital control from $D_0$ to $D_{N-1}$ which corresponds to a digital input from 1 to $2^N-1$, the width or the number of fins can be discretized across multiple transistors in parallel (wire-ORed i.e. all of the drain terminals are tied together) driving a common MTJ. The effective width of the transistor corresponding to a more significant bit 'm' would be twice the width or twice the number of fins of the transistor corresponding to the adjacent less significant bit 'm-1'. In the DCO design, the gate terminal receives full scale voltage as logic 1 and the transistor is turned-off for logic 0 on receiving 0 V input; hence magnitudes of both its $V_{GS}$ and $V_{DD}$ are fixed. This scheme enables an electronic control of the effective channel width or an effective transistor whose channel-width directly corresponds to the digital input. The lowest digital input of 1 obtained by setting only lowest significant bit to logic 1 and others to logic 0 would correspond to the lowest targeted frequency of the oscillator, while the input $2^N-1$ obtained by setting all of the N-bits to logic 1 state would correspond to the largest targeted frequency, where the ratio of the largest to the lowest frequency would give us the tunability ratio for the DCO.

Figure 14(b) shows one such example of a 3-bit coarse DCO for 65 nm CMOS node operating with both $V_{DD}$ and $V_{GS}$ at 1.1 V, with gate input being controlled digitally as either 0 V (bit 0) or 1.1 V (bit 1). The channel width of three transistors, $W_0$, $W_1$ and $W_2$, is respectively designed to be 7 L, 14 L and 28 L, where L is the channel-length. The digital input vector [$D_2$ $D_1$ $D_0$] results in 7 possible digital controls 001, 010, 011, 100, 110 and 111, not counting 000 as it turns-off all the transistors and switches off the oscillator. These controls correspond to an effective W/L ratio of 7, 14, 21, 28, 35, 42 and 49. This discretizes the frequency trend of Fig. 13(c) (blue circles) into 7 output frequencies from 119 MHz to 334 MHz, with a tuning ratio of 2.8. The DCO output characteristic in Fig. 14(b) is observed to be saturating, which implies non-linearity in the input-to-output mapping of the digital-control-to-frequency over the designed frequency regime, in-line with the behavior of the current shown in Fig. 13(a). For a larger degree of linearity, DCO can also be re-designed with the largest W/L, corresponding to $2^N-1$ input, restricted to be around 20, but it would result in a maximum frequency of 271 MHz for the chosen set of NMOS and MTJ parameters.

## IV. SUMMARY

We propose a fundamentally new concept of obtaining oscillations with frequencies in VHF and UHF bands. We posit a novel scheme of the spin-torque oscillators which is based on the superior full-switching of the free-layers in contrast to the extant precessional orbit-based schemes. Moreover, it is shown that the free-layers can be reversibly switched without reversing the direction of the current. Oscillator is shown to operate without any external magnetic field and any pinned layer. Absence of the need of a pinning layer may in fact result in a smaller stack height and integration of the oscillator among lower metallic layers in contrast to standard MTJs which are pushed to higher layers because of a thick stack-size. The oscillator can operate in the range of 48 MHz to 965 MHz, with the tuning ratio of 13.6 (CCO) and 5.8 (VCO), output from −26 to −48 dBm. However, presently, the quality-factor (Q) is in the same range as for free-running STNO and thus need a lot of improvement to come on par with much higher Q, in several hundreds to thousands, indulged by the precisely biased STNO and/or the synchronized/locked oscillators. Furthermore, although the power-efficiency of a SW-STO in this work is better than that of the STNOs for low frequencies, it becomes much worse on approaching 1 GHz and remains much inferior to the ring oscillators throughout the investigated range. Hence, power efficiency is another performance metric, besides the Q factor and the output power, which needs further work to make SW-STO truly a competitive solution for the tunable oscillators. Next, the analysis including the driving NMOS shows that the NMOS dominates the design area and the larger nodes would be better for driving the proposed oscillator. Moreover, because of the dominance of the transistor on the area, the scaling constraints on the MTJ can be relatively relaxed, and the oscillator performance can be further improved compared to results generally presented in this work. VCO configuration can furthermore be extended into a digital design and used as a DCO. Table I illustrates some of our exemplar results (second and third column) against other oscillators in the literature. It shows that our design may indeed enable a very large tuning ratio, with a modest Q factor and a very small area for a decent output signal and power-efficiency in the frequency range useful for ASIC clocks and long-range low-power IoT communication via LoRa protocol.

**Corresponding Author**

[†]gauravdce07@gmail.com

**ACKNOWLEDGEMENTS**

This work at the National University of Singapore was supported by CRP award no. NRF-CRP12-2013-01 and MOE2013-T2-2-125. We gratefully acknowledge the discussions with Xuanyao Fong, the useful review comments on the manuscript from Kien Trinh Quang and Sachin for helping with Fig. 2.

**Figures**

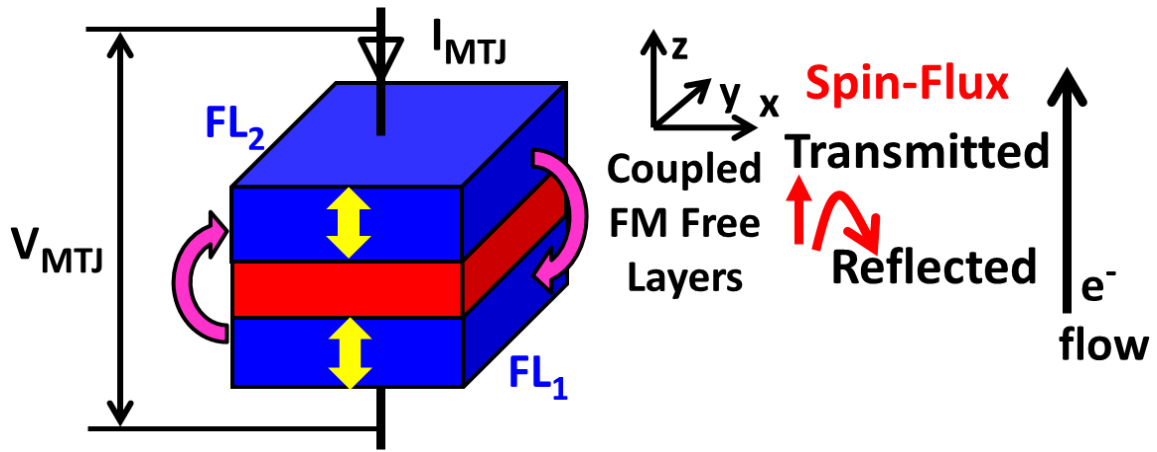

**Figure 1**. Conceptual illustration of a pMTJ with a dipolar coupled dual free-layer, without any pinned layer. It is driven by a uni-directional current from $FL_2$ to $FL_1$. The transmitted spin-flux with a vector parallel to the magnetization of $FL_1$ acts on $FL_2$, while reflected spin-flux with a vector anti-parallel to the magnetization of $FL_2$ acts on $FL_1$, which for a set of electronic and geometric constraints results in a self-sustained switching of both the free layers producing AC oscillations for a DC bias. When this pMTJ is driven via tunable DC current source $I_{MTJ}$, the device is in a CCO configuration.

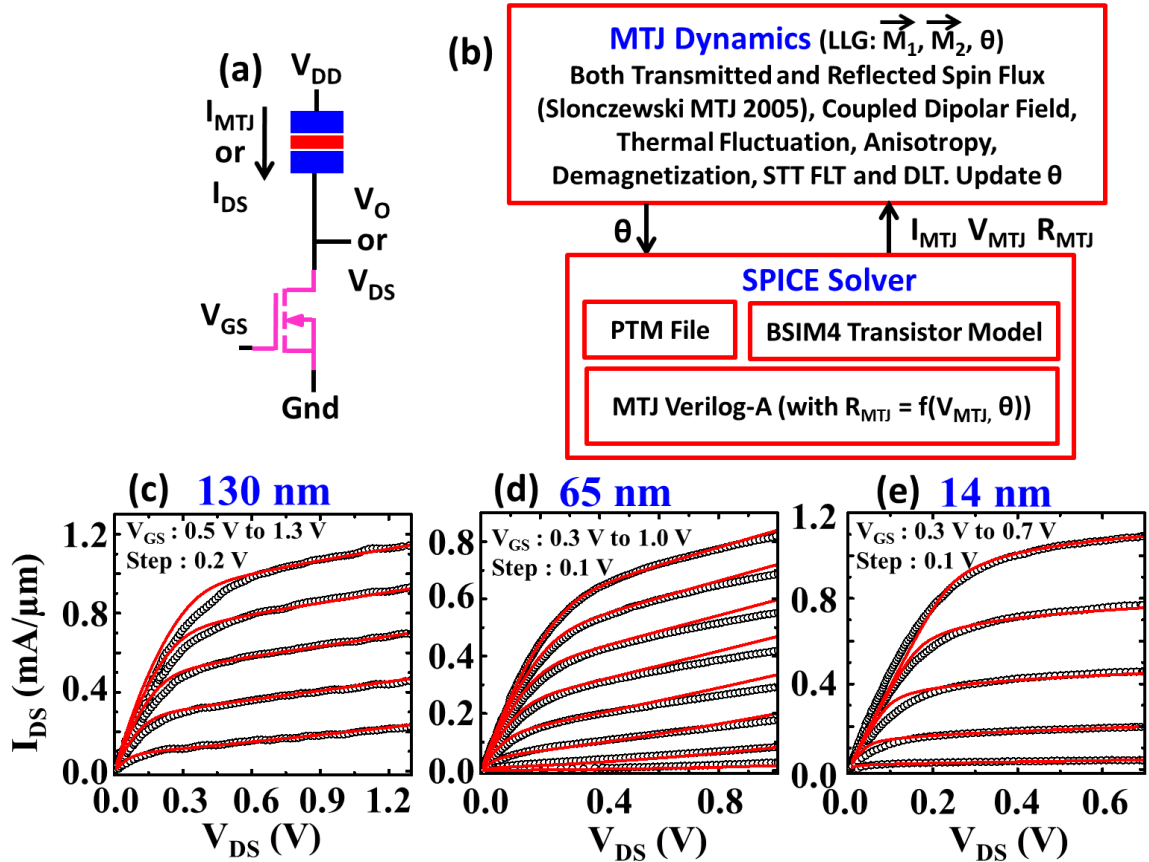

**Figure 2**. (a) VCO configuration with a pMTJ driven via tunable DC $V_{GS}$ for a fixed $V_{DD}$ (gate-control) or tunable DC $V_{DD}$ for a fixed $V_{GS}$ (drain-control). (b) Schematic summary of the simulation framework expounded in the methodology section. (c-e) Published experimental data (black circles) of the NMOS output characteristics fitted (red line) against the BSIM4 model used in a spice solver for three representative CMOS nodes. For (c) 130 nm node fitting against Fig. 4 of Ref. [76] from Intel, (d) 65 nm node fitting against Fig. 6 of Ref. [77] from IBM, Chartered and Infineon, and (e) 14 nm node fitting against Fig. 5 of Ref. [78] from Intel.

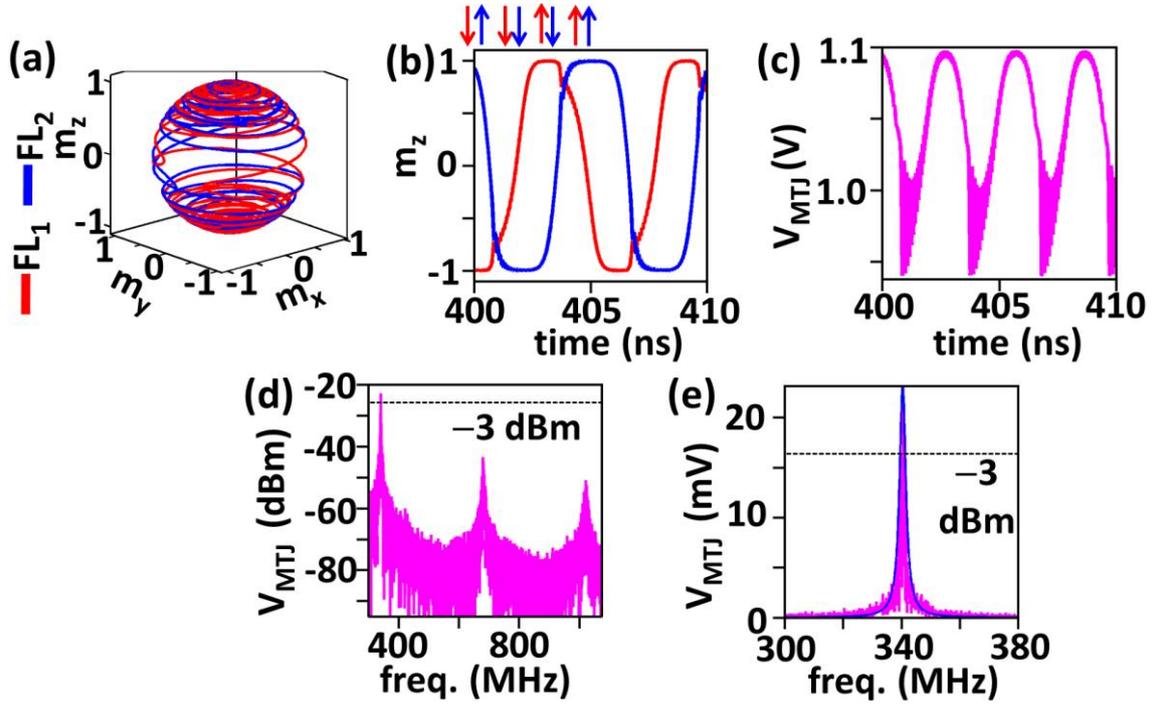

**Figure 3**. Operation principle (without a thermal field) of the device driven by a constant current. (a) 400-405 ns dynamics of the magnetization unit vector **m** = [$m_x$ $m_y$ $m_z$]$^T$. (b) Full-switching of both $m_z$. (c) Output voltage oscillations $V_{MTJ}$. FFT of $V_{MTJ}$ in dBm (d) and mV (e) demonstrating a fundamental tone along with the first two harmonics (in (d)) and the Lorentzian fit (blue line in (e)). The dashed black lines are −3dBm line which cuts through half-power points. Lorentzian fit is used to extract the linewidth and the quality factor of the oscillator.

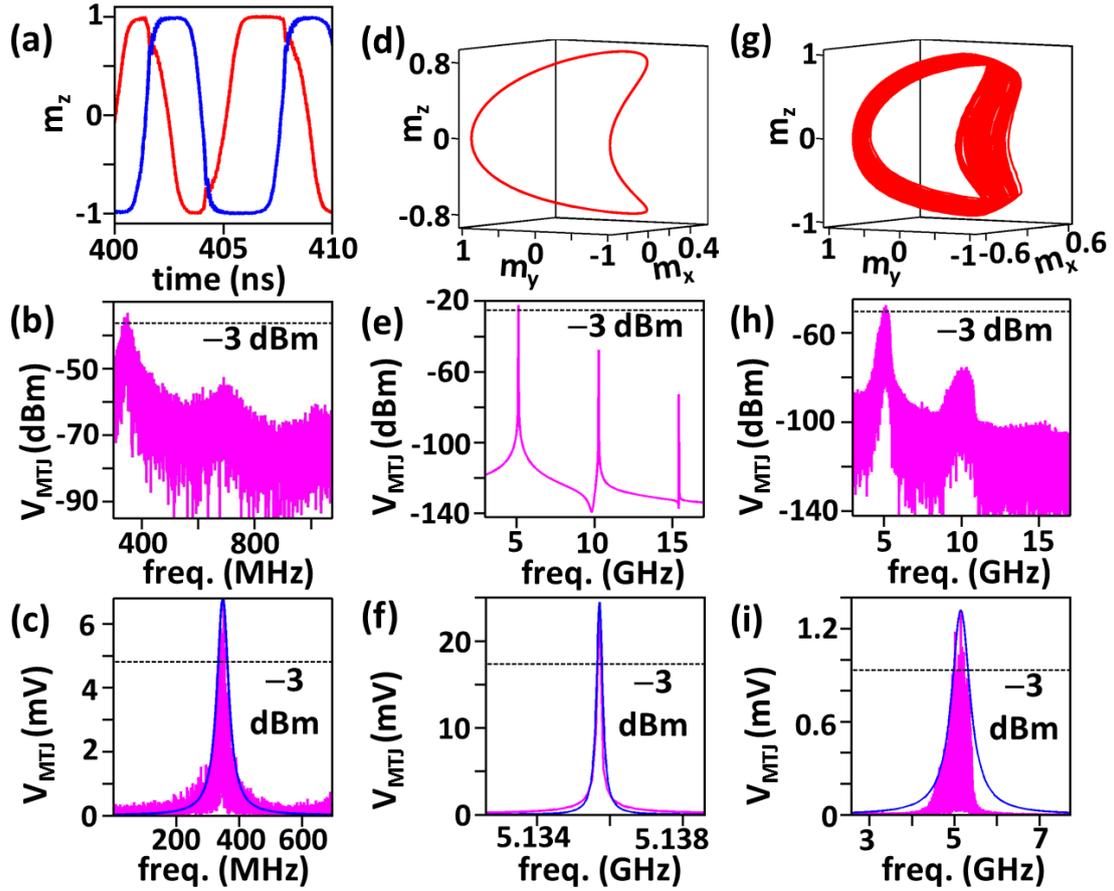

**Figure 4.** Effect of the thermal fluctuation field (300 K) on the proposed oscillator (a-c) is contrasted with an orthogonal STNO (c-i) of 60nm×40nm×5nm(1.8nm) with an in-plane fixed layer (in place of $FL_1$) and a perpendicular free layer (in place of $FL_2$) operating at 5.1 GHz with 70 μA fixed current without an external bias field. (c-f) STNO without a thermal field clearly shows the precessional orbit (d) and FFT of the voltage across the MTJ (e, f). (g-i) STNO with a thermal field shows randomness introduced in the orbit ((d, g) illustration represents 400-450 ns of precession), and the linewidth broadening in FFT in (h, i). In FFT plots, the dashed black lines are −3dBm line which cuts through half-power points and thus determines the linewidth and the quality factor of the oscillator, while the blue lines are the Lorentzian fit via which the linewidth and the quality factor of the oscillator are extracted.

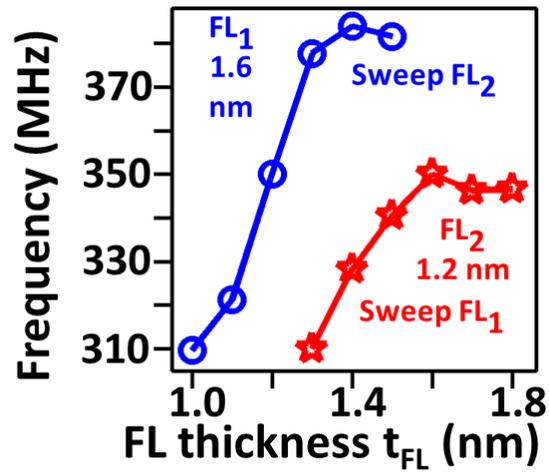

**Figure 5**. Effect of the FL thickness on the oscillation frequency for a representative $I_0$ of 500 µA. MTJ can thus be designed to meet the frequency specifications of an oscillator over a wide frequency range.

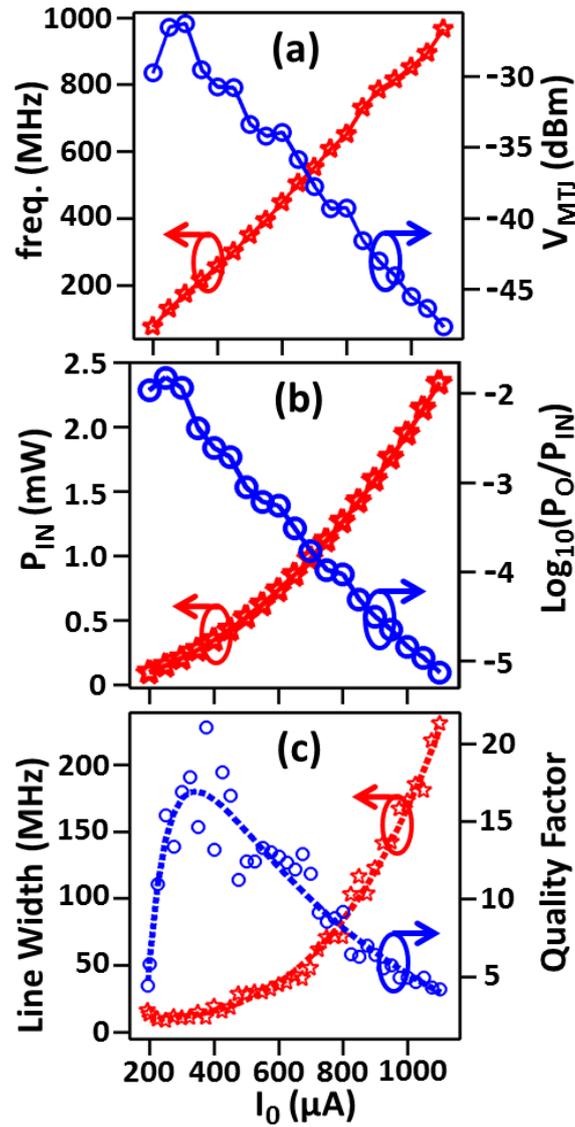

**Figure 6.** Effect of a constant current $I_0$ on (a) the frequency and the output voltage oscillations, (b) the input power consumed $P_{IN}$ and the power efficiency ($P_O/P_{IN}$) i.e. the delivered output power to the consumed power, (c) the linewidth and the quality factor (Q) of the oscillator. Dashed lines are the fit to the data to exhibit an overall trend. (b) Consumed power decreases directly with the current thereby improving the power efficiency especially at the lower frequencies. (c) Linewidth (frequency spectra) narrows at the smaller currents (frequencies) which improve the quality factor (= Center Frequency / Linewidth) at intermediate frequencies up to 21.

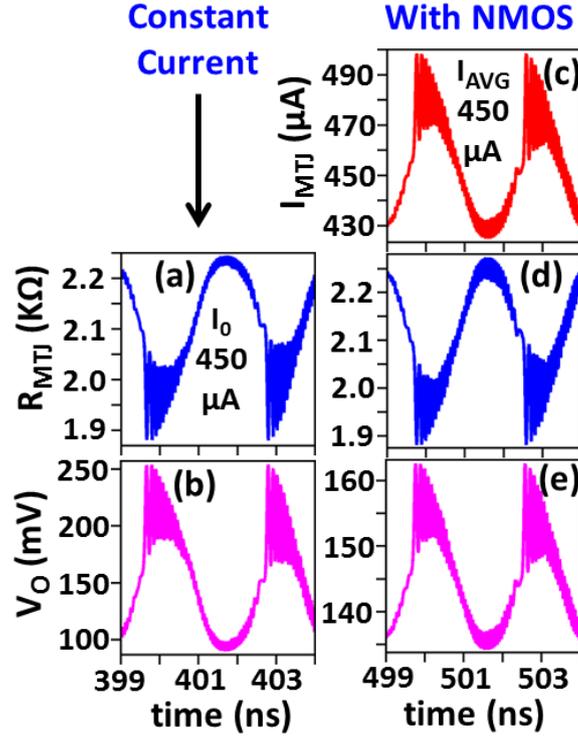

**Figure. 7.** Effect of replacing (a, b) a constant current source with (c-e) an NMOS (W/L=28) to drive the MTJ. For (b) $V_O = 1.1 - V_{MTJ}$ for an appropriate comparison with (e). For an NMOS, the drain current also oscillates in anti-phase with the resistance, thereby reducing the output swing which is approximately given by eq. (10). $I_0$ is chosen equal to the average drain current, and the complementary MTJ voltage is used as $V_O$ for a constant current case, for the rightful comparison.

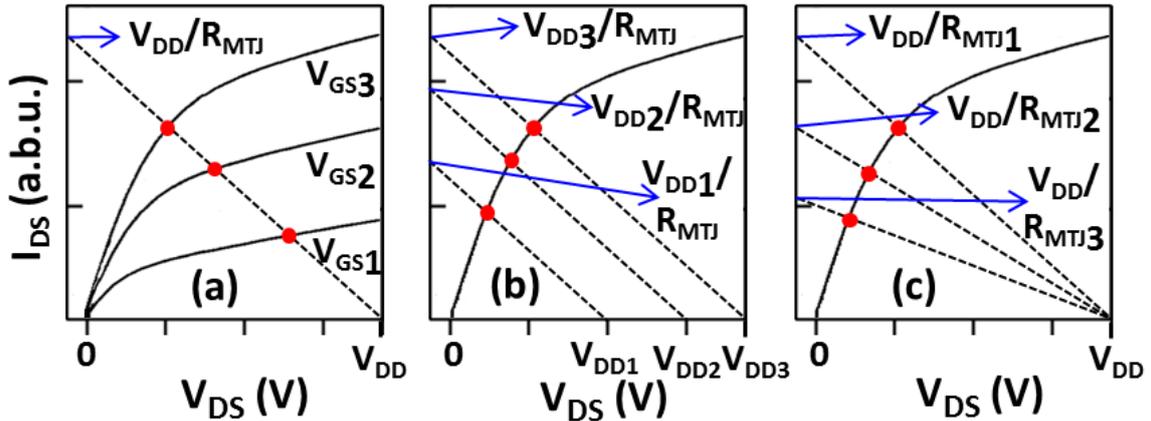

**Figure 8.** Conceptual illustration of load-line analysis to show the effect of sweeping (a) $V_{GS}$, (b) $V_{DD}$ and (c) MTJ cross-section. Larger index in the subscript represents a larger magnitude of the corresponding quantity. Solid black line is the NMOS output characteristics. Dashed black line is the $R_{MTJ}$ output characteristic, where $R_{MTJ}$ is treated as a constant trivial resistor in this illustration. Quiescent point is marked via red dot at the point of an intersection of the $R_{MTJ}$ and the NMOS characteristics. This serves as a visual aid for understanding the data for the effect of the respective parameters shown in the subsequent figures.

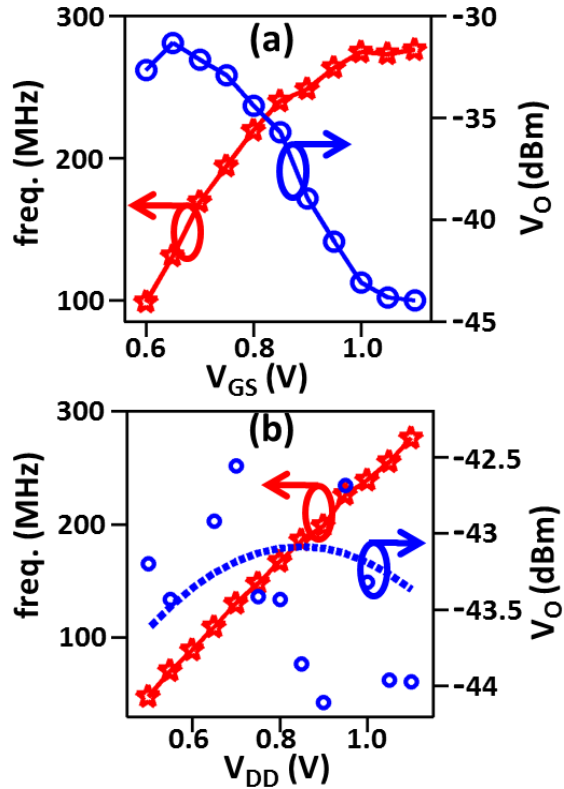

**Figure 9.** Effect of sweeping (a) gate voltage $V_{GS}$ and (b) node voltage $V_{DD}$ on frequency and output voltage $V_O$. Dashed lines in (b) is the fit to the data to exhibit an overall trend for $V_O$.

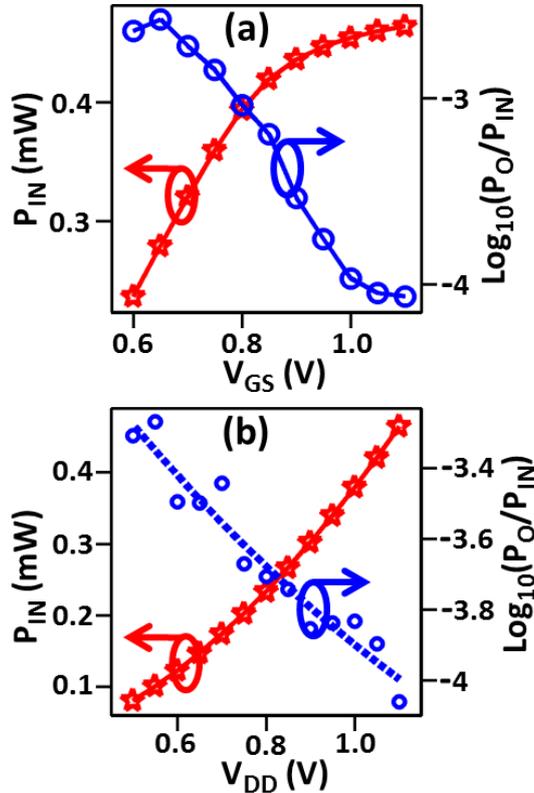

**Figure 10.** Effect of sweeping (a) gate voltage $V_{GS}$ and (b) node voltage $V_{DD}$ on consumed input power $P_{IN}$ and power efficiency $P_O/P_{IN}$. Dashed line in (b) is the fit to the data to exhibit an overall trend for $P_O/P_{IN}$.

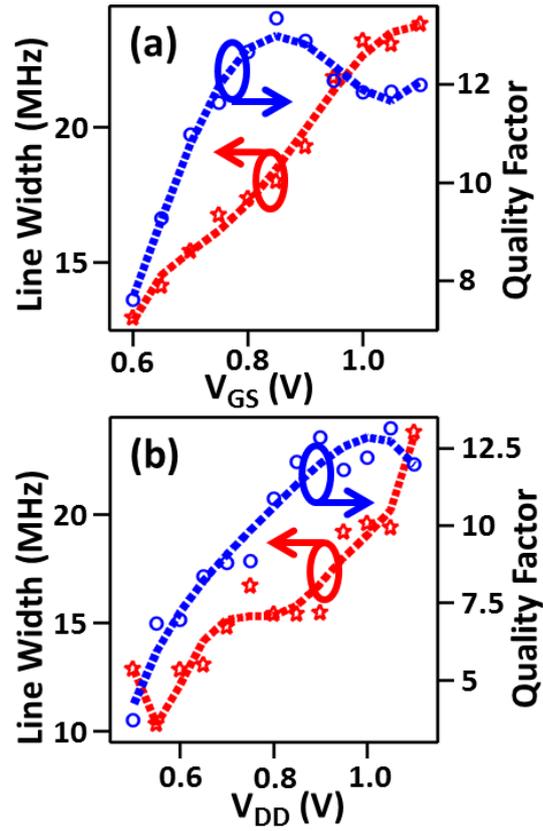

**Figure 11.** Effect of sweeping (a) gate voltage $V_{GS}$ and (b) node voltage $V_{DD}$ on Linewidth and quality factor Q. Dashed lines in are the fit to the data to exhibit an overall trend.

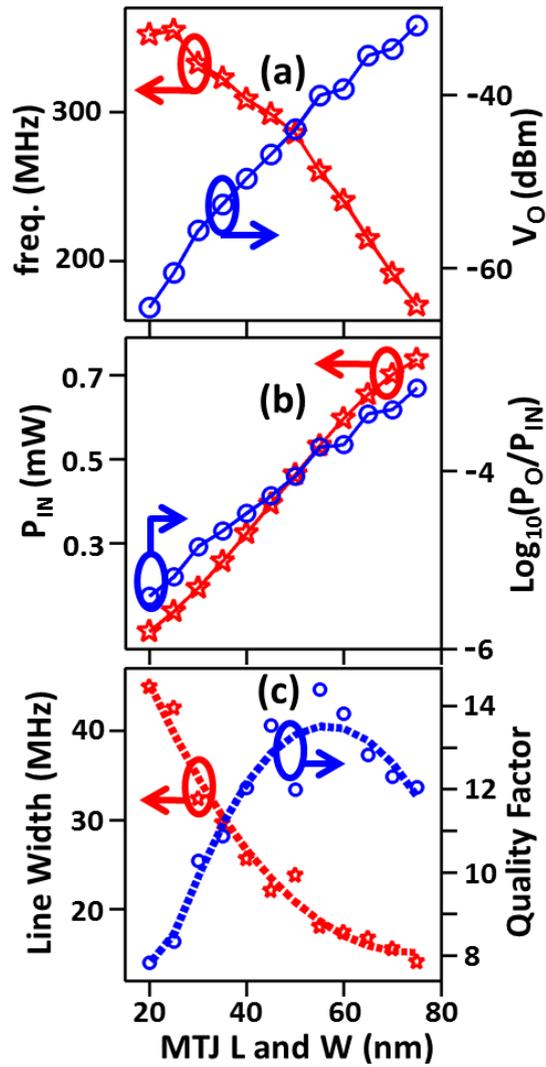

**Figure 12.** Effect of sweeping MTJ cross-section (length=width) on (a) frequency and output voltage $V_O$, (b) input power $P_{IN}$ and power efficiency $P_O/P_{IN}$, and (c) Linewidth and quality factor Q. Dashed lines in (c) are the fit to the data to exhibit an overall trend.

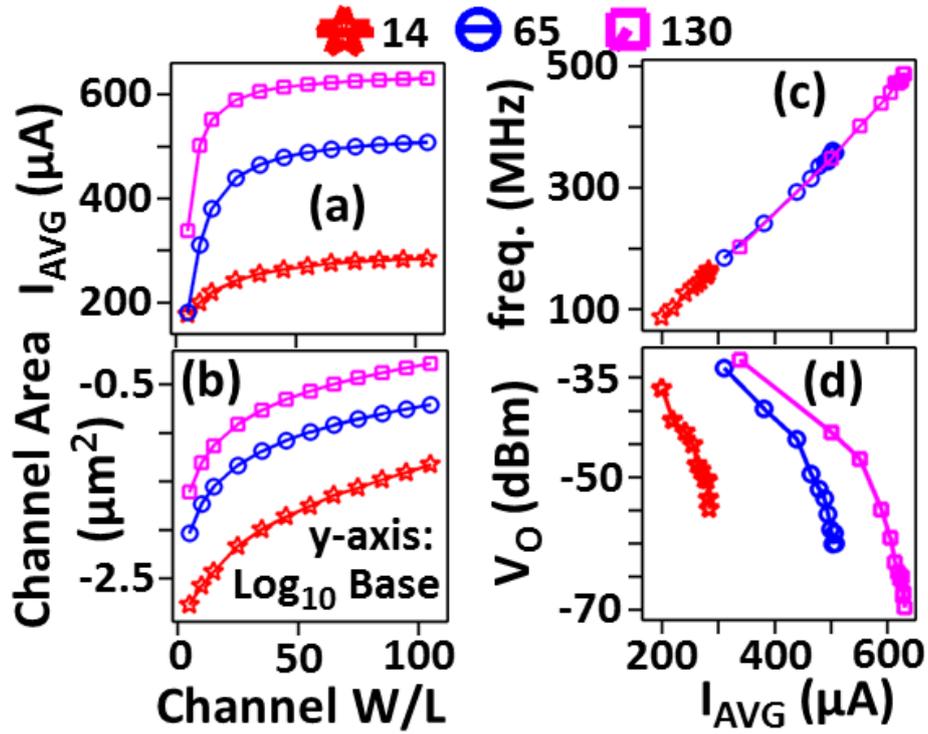

**Figure 13**. Effect of W/L of transistors on performance across 14 nm, 65 nm and 130 nm node operating at the full scale node voltage (0.7, 1.1, 1.3 V) i.e. $V_{DD}=V_{GS}$. (a) Average operating current for the oscillator. (b) Channel area for the reference. Output frequency (c) and the signal amplitude (d).

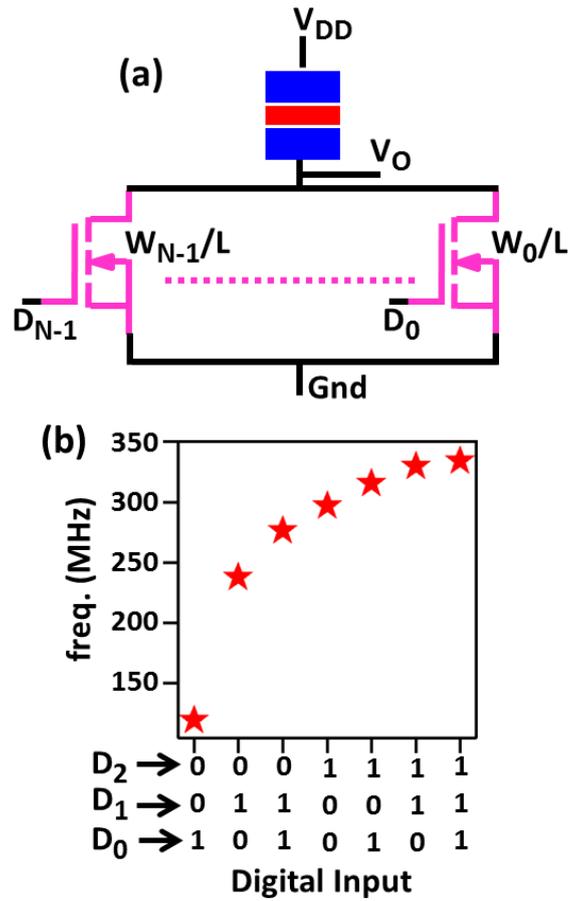

**Figure 14**. (a) Re-designing a VCO into a DCO configuration by splitting the effective transistor width across wire-ORed N-transistors, driven by an N-bit digital control $D_0$ to $D_{N-1}$. Effective width for a bulk-NMOS or number of fins for a FinFET of transistor 'm', driven by a bit $D_m$, is twice the width or number of fins of a transistor 'm-1' where m∈[1,N-1]. (b) Frequency obtained for a 3-bit DCO with channel widths of $W_2 = 2\ W_1 = 4\ W_0 = 28\ L$, with a tuning ratio of 2.8.

Table I. Summary of a comparison with other designs in the literature.

| Criteria | pp ϑ (C) | pp ϑ, β (NMOS) | STNO κ, χ, ƌ (pi/ip) (C) | Ring λ, ζ [42] |
|---|---|---|---|---|
| Freq. (f) (MHz) | 71 – 965 | 47.6 – 276.3 | 600 – 4000 [20, 41, 79, 80] | 480-1100 |
| $f_{max}/f_{min}$ | 13.6 | 5.8 | ~ 2 & | 1.25 |
| $V_O$ (dBm) | −26 to −48 | −31 to −44 | −42 & | −2.97 |
| $P_O/P_{IN}$ (%) | 1.5 to 0.0007 | 0.26 to 0.008 | 0.12 & | 13.1 |
| Q | 4.2 - 21 | 3.7 – 13.3 | 8-39 & | −120 to −108 dBc/Hz Φ |
| Area (µm$^2$) | 0.0025 * | 0.0388 # | 0.0105 *& | 14784 Ψ |

C: Driven by Current Source; * MTJ area; # Channel Area; Ψ Core Area; ϑ. 50×50×1.6(1.2); β. Gate Control 65 Node W/L=21; δ. 50×40×t1(t2); κ. ii and pp need an external B-Field to precesses. χ. TMR based without any additional locking or post-processing circuit; ƌ See Table 1 of Ref. [41]; & Best of the results of Ref. [41]; λ LC is not compared, because in the sub-GHz range the size of an on-chip inductor and a capacitor is in hundreds of µm$^2$. Its comparison is better suited in high GHz range. Φ Phase Noise (not Q); ζ with PLL lock (not for free-running case).